\documentclass[a4paper, 11pt]{article}

\usepackage[switch]{lineno}
\modulolinenumbers[5]
\usepackage{xcolor}
\usepackage{placeins}
\usepackage[utf8]{inputenc}
\usepackage[colorlinks, linkcolor = black, citecolor = black, filecolor = blue, urlcolor = blue]{hyperref}
\usepackage{amsmath,amssymb}
\usepackage{graphicx}

\newcommand{\mySphereWidth}{0.48}
\newcommand{\mySkymapWidth}{0.7}
\definecolor{sgdarkgreen}{rgb}{0.2,0.7,0.05}
\definecolor{sgred}{rgb}{1.0,0.1,0.1}

\usepackage{authblk}
\begin{document}

\title{Model-independent search for neutrino sources\\ with the ANTARES neutrino telescope}

\author[b]{A.~Albert}
\author[c]{M.~Andr\'e}
\author[d]{M.~Anghinolfi}
\author[e]{G.~Anton}
\author[f]{M.~Ardid}
\author[g]{J.-J.~Aubert}
\author[h]{T.~Avgitas}
\author[h]{B.~Baret}
\author[i]{J.~Barrios-Mart\'{\i}}
\author[j]{S.~Basa}

% ...
\affil[b]{\scriptsize{GRPHE - Universit\'e de Haute Alsace - Institut universitaire de technologie de Colmar, 34 rue du Grillenbreit BP 50568 - 68008 Colmar, France}}
\affil[c]{\scriptsize{Technical University of Catalonia, Laboratory of Applied Bioacoustics, Rambla Exposici\'o,08800 Vilanova i la Geltr\'u,Barcelona, Spain}}
\affil[d]{\scriptsize{INFN - Sezione di Genova, Via Dodecaneso 33, 16146 Genova, Italy}}
\affil[e]{\scriptsize{Friedrich-Alexander-Universit\"at Erlangen-N\"urnberg, Erlangen Centre for Astroparticle Physics, Erwin-Rommel-Str. 1, 91058 Erlangen, Germany}}
\affil[f]{\scriptsize{Institut d'Investigaci\'o per a la Gesti\'o Integrada de les Zones Costaneres (IGIC) - Universitat Polit\`ecnica de Val\`encia. C/  Paranimf 1 , 46730 Gandia, Spain.}}
\affil[g]{\scriptsize{Aix-Marseille Universit\'e, CNRS/IN2P3, CPPM UMR 7346, 13288 Marseille, France}}
\affil[h]{\scriptsize{APC, Universit\'e Paris Diderot, CNRS/IN2P3, CEA/IRFU, Observatoire de Paris, Sorbonne Paris Cit\'e, 75205 Paris, France}}
\affil[i]{\scriptsize{IFIC - Instituto de F\'isica Corpuscular (CSIC - Universitat de Val\`encia) c/ Catedr\'atico Jos\'e Beltr\'an, 2 E-46980 Paterna, Valencia, Spain}}
\affil[j]{\scriptsize{LAM - Laboratoire d'Astrophysique de Marseille, P\^ole de l'\'Etoile Site de Ch\^ateau-Gombert, rue Fr\'ed\'eric Joliot-Curie 38,  13388 Marseille Cedex 13, France}}
% ...

\date{April 14, 2019}

\maketitle
\bigskip

\begin{abstract}
\noindent
A novel method to analyse the spatial distribution of neutrino candidates
recorded with the ANTARES neutrino telescope is introduced, searching for an
excess of neutrinos in a region of arbitrary size and shape from any direction
in the sky. Techniques originating from the domains of machine learning, pattern
recognition and image processing are used to purify the sample of neutrino
candidates and for the analysis of the obtained skymap. In contrast to a
dedicated search for a specific neutrino emission model, this approach is
sensitive to a wide range of possible morphologies of potential sources of
high-energy neutrino emission. The application of these methods to ANTARES data
yields a large-scale excess with a post-trial significance of 2.5$\sigma$. 
Applied to public data from IceCube in its IC40 configuration, an excess
consistent with the results from ANTARES is observed with a post-trial
significance of 2.1$\sigma$.
\end{abstract}


\section{Introduction}

Since the recent discovery of a diffuse high-energy astrophysical neutrino flux
by the IceCube Collaboration
\cite{Aartsen:2013bka,Aartsen:2013jdh,Aartsen:2014muf}, neutrino astronomy has
established itself as a new discipline. Due to the statistical limitations
implied by a new observational tool that has just overcome its initial detection
threshold, the spectral and spatial properties of the discovered flux are still
not well constrained.

The high-energy starting event (HESE) analysis \cite{Aartsen:2013jdh}, which is
most sensitive to the Southern sky but has a poor spatial resolution, measures a
best-fit diffuse signal of $dN_{\nu}/dE \propto E^{\Gamma}$ with $\Gamma = -2.5$
\cite{Aartsen:2015knd}. The flux observed from the Northern Sky with a higher
energy threshold of about 200\,TeV in the muon channel exhibits a harder
spectral index of about $\Gamma=-2.13\pm0.13$ \cite{Aartsen:2015rwaUpdate2016}.

A first analysis has shown that the spatial distribution of HESE events is
consistent with an isotropic distribution \cite{Aartsen:2014gkd}.

While there seems to be some evidence for an excess at low galactic latitudes
\cite{Neronov:2015osa}, favouring a galactic contribution with a softer spectrum
and an extragalactic contribution with a harder spectrum \cite{Neronov:2016bnp},
the origin of the various contributions remains open.

Several analyses with the goal to reveal the origin of the astrophysical
neutrinos have been performed. Time-integrated searches for point-like and
extended bright sources by ANTARES \cite{Adrian-Martinez:2014wzf}, IceCube
\cite{Aartsen:2014cvaUpdate2017} as well as a joint search
\cite{Adrian-Martinez:2015ver} exclude the possibility that the flux can be
generated by a small number of bright sources. A first search employing a
two-point correlation function with data from the ANTARES neutrino telescope
\cite{Adrian-Martinez:2014hmp} has neither found significant deviations from
isotropy in the full-sky neutrino distribution nor shown evidence for
correlations with catalogues of various astrophysical objects. Likewise a
two-point correlation and multipole analysis of IceCube skymaps confirmed that
the assumption of a small number of bright sources is excluded
\cite{Aartsen:2014ivk}. Therefore, a distribution of many faint point-like or
unexpected extended sources constitutes a promising hypothesis at this stage.

ANTARES \cite{antares:antares} is the largest operational neutrino telescope in
the Northern Hemisphere, located in the Mediterranean Sea (42$^\circ$48'N,
6$^\circ$10'E) at a depth of 2475\,m. Due to its location, it mainly observes
the Southern sky in the upgoing muon channel and provides an excellent view of
the Galactic Centre. Despite its much smaller instrumented volume compared to
IceCube, it has an effective area for muon neutrinos which is comparable to that
of the IceCube HESE analysis for energies around 100\,TeV and even surpasses it
for energies below about 60\,TeV \cite{Adrian-Martinez:2015jlj}.

This paper introduces three independent new methods and the results obtained
with them. The first two algorithms enhance the data selection and
reconstruction process, while the third is a novel analysis method, referred to
as {\it multiscale method} in the following, that uses the arrival directions
of upgoing muon neutrinos recorded with the ANTARES neutrino telescope in 6
years of data taking.

The goal of this analysis is to identify the most significant, spatially
confined excess over the background of atmospheric neutrinos without relying on
assumptions of any emission model of potential neutrino sources. The analysis
method is sensitive to a clustering of events in an extended sky region, and its
sensitivity increases with the size of the signal region. The multiscale method
does not rely on any property derived from modelling or simulations, but
computes all required estimates from the observed data. Compared to other
searches for potential extended sources, it has the benefit of not being
restricted to a given template. In contrast to other model-independent searches,
like e.g.\ a two-point correlation analysis, it identifies the region of a dense
clustering. The method is designed to yield a first indication of a candidate
source region significantly deviating from the expected background distribution,
but not to provide a detection with maximum significance or to interpret the
nature of a candidate cluster.

\section{Data selection and reconstruction}
\label{sec:dataSelectionAndReco}

\subsection{Signal identification}
\label{sec:updown}

In order to distinguish events resulting from genuine neutrinos from the
background of atmospheric muons that reach the telescope from above and generate
about $10^6$ times more events, cuts are placed on the direction reconstruction
to select those that are consistent with an ``upgoing'' particle entering the
telescope from below.

\begin{figure}[ht]
\begin{center}
  \includegraphics[width=0.5\textwidth]{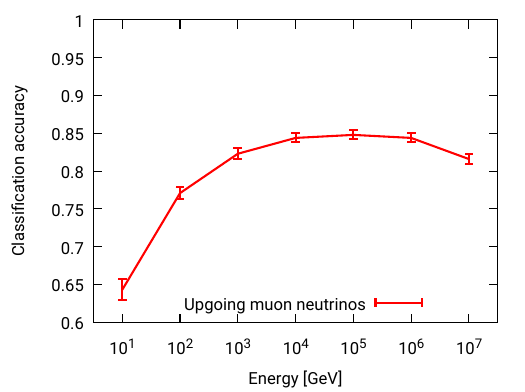}
\end{center}
\caption{%
The classification accuracy of the two step RDF classification for upgoing muon
neutrinos versus neutrino energy for Monte Carlo simulations. The error bars
indicate one standard deviation of statistical errors plus an estimate of the
systematic error resulting from fluctuations in the training sample.}
\label{fig:updownacc}
\end{figure}

In this analysis a multivariate classification technique called ``Random
Decision Forest'' (RDF)\footnote{The used implementation is forked from an open
source version of alglib \cite{alglib:website}.} \cite{ho:RDFs} is used in
addition to cuts on the reconstruction quality to allow for less strict cuts,
increasing the available statistics. The RDF operates on variables describing
the topology and timing pattern of the light observed within ANTARES
\cite{geisselsoeder:phdARXIV}. The output of this algorithm for a recorded event
is an assignment to a predefined class. In this application the classes are
``upgoing'' and ``downgoing''. The algorithm is trained on Monte Carlo
simulations that incorporate the observed time-dependent data-taking conditions
\cite{antares:rbr}. To improve the accuracy of the results, a two-step
classification is used, where the first classification rejects only clearly
downgoing signatures, while the second step is trained specifically to filter
out those atmospheric muon events that generate patterns similar to the desired
upgoing muon neutrinos. This technique reaches a rejection rate of $99.85\% \pm
0.12\%$ for downgoing atmospheric muons while preserving $81.7\% \pm 1.3\%$ of
all charged current muon neutrino events (all numbers with respect to all
triggered events, calculated for a spectrum following $E^{-2.0}$). Compared to a
single stage RDF classification with a similar efficiency for upgoing neutrino
events it reduces atmospheric muons by a factor of 20. 
Figure~\ref{fig:updownacc} shows the efficiency for upgoing muon neutrinos as a
function of the neutrino energy. The classification accuracy is defined as
$\frac{\textrm{Number of correctly classified events}}{\textrm{Total number of
events}}$.

\clearpage

\subsection{Direction reconstruction}
\label{sec:selectfit}

A novel method called ``Selectfit" is used to reconstruct the direction of
neutrino candidates. Instead of applying one reconstruction algorithm for all
neutrino candidates, Selectfit combines the results of multiple direction
reconstruction algorithms, aiming to select the most precise result for each
event. It combines four reconstruction algorithms previously used by ANTARES
\cite{antares:psandaafit,antares:bbfit,erwin:phdARXIV}. While the reconstruction
schemes of the algorithms that are combined are similar, each algorithm performs
best for a different energy range or has different prerequisites for a
successful reconstruction, allowing Selectfit to improve the overall result. The
selection is again performed by a Random Decision Forest using the
reconstruction results (zenith and azimuth angle) as well as all available
quality-related output parameters of the considered reconstruction algorithms as
input variables. It tries to identify the most accurate reconstruction result. 
The output class number hence determines which algorithm is used for an event.

In order to estimate the accuracy of the reconstruction for each event in a
comparable way, Selectfit cannot just rely on the quality variable of the chosen
algorithm. Instead it combines the results and quality parameters of all
considered algorithms using a RDF. The output of this second RDF are classes
corresponding to bins of angular error ( $<0.1^{\circ}$, $<0.2^{\circ}$,
$<0.4^{\circ}$, $<0.8^{\circ}$, \dots).

\begin{figure}[ht]
 \begin{center}
   \includegraphics[width=0.5\textwidth]{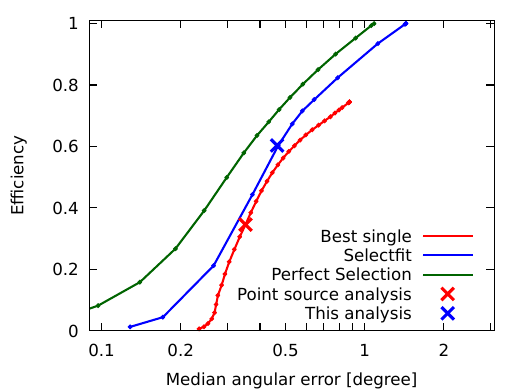}
\caption{%
Comparison of the efficiency versus the median angular error of direction
reconstruction algorithms. Every point of these curves is obtained for a
different cut on the quality variables. Shown are the performance of the on
average most accurate individual direction reconstruction algorithm
\cite{antares:psandaafit} in red for cuts on its standard quality parameter and
the novel method ''Selectfit`` with cuts on the estimated error class in blue. 
The crosses indicate the cuts used for the single algorithm in \cite{antares:ps}
and for Selectfit in this analysis. The green line shows the unreachable limit
for the perfect combination of direction reconstruction algorithms. It is the
performance if for every event the best available single reconstruction was
chosen and the true angular uncertainty was used for the selection. All
performances have been evaluated for a neutrino flux following an $E^{-2}$
energy spectrum. Efficiency is defined with respect to all triggered events
where at least one direction reconstruction succeeded.}
\label{fig:selectfit}
\end{center}
\end{figure}

As illustrated in Figure~\ref{fig:selectfit}, Selectfit either allows the
angular error for a sample of neutrinos to be reduced or it can be used to
increase the sample size for a given angular error. It increases the available
statistics by at least 11\% for any given accuracy.
The benefit for small angular errors is mainly due to the improved estimation of
the angular error, whereas for less accurate reconstructions, the main benefit
results from the event-wise selection from the four reconstruction methods
\cite{geisselsoeder:phdARXIV}.

\begin{figure}[ht]
 \begin{center}
   \includegraphics[width=0.5\textwidth]{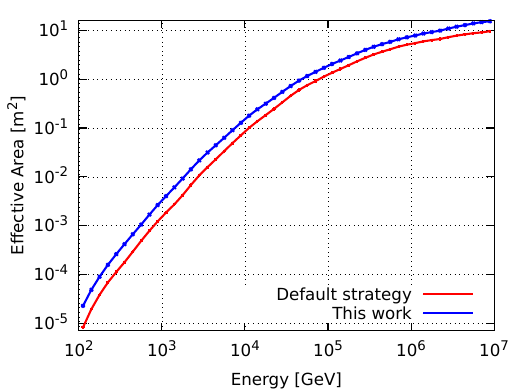}
\caption{%
Comparison of the effective area for the default event selection and
reconstruction strategy for point sources \cite{antares:ps} in red and this work
in blue. The ratio is about 2.5 at 100\,GeV and decreases to about 1.5 for
energies above 100\,TeV.}
\label{fig:effA}
\end{center}
\end{figure}

Since a search for extended sources does not necessarily require the same
angular precision as a point source search, a looser cut on the uncertainty of
the reconstruction is applied. For a flux of muon neutrinos with an energy
spectrum of $E^{-2}$ these cuts result in a median angular uncertainty of
0.46$^{\circ}$. The cut for this analysis was chosen to yield a high statistics
sample of muon neutrino candidates while reconstructing them accurately enough
to match the 0.5$^{\circ}$ grid spacing used in the multiscale method discussed
in the following section. In Figure \ref{fig:effA} the effective area reached
with this method is compared to the effective area obtained with the default
reconstruction scheme as used for point-source searches \cite{antares:ps}. The
looser event selection in this analysis also results in a higher background from
misreconstructed atmospheric muons in the final sample of neutrino candidates
(29.7\% instead of 10\% in \cite{Adrian-Martinez:2014wzf}).

\clearpage

\section{Model-independent multiscale source search}
\label{sec:multiscale}

The model-independent multiscale source search aims to identify the region of
arbitrary position, size, shape and and distribution of neutrino events within
the region with the most significant excess of neutrino events in the sky with
respect to the background expectation.

The method itself\footnote{The source code of a slightly altered version of this
method can be found at \url{https://github.com/sgeisselsoeder/multiscale} and
the method can easily be tested using
\url{https://hub.docker.com/r/km3net/multiscale}.} is independent from the
ANTARES neutrino telescope or the fact that the analysed data are neutrino
candidates. The only assumption made in this method is that the density of
background events at any declination is approximately independent of the right
ascension. Unlike many techniques known from image processing, it is designed to
operate also on the sparse dataset of neutrinos. While many other fields of
astronomy can operate on image-like measurements, these neutrino data would fill
only very few grid points of a skymap. Hence many methods successfully used in
other experiments are not applicable here.

The search method consists of eight subsequent steps applied to the measured
neutrino sky map, described in subsections 3.1--3.8. These steps have been
defined such that they are independent from assumptions on signal topology or
spectrum. They implement measures to suppress background and reduce
fluctuations, to identify regions that stick out from the background, and to
determine topology and significance of a possible signal. They are the result of
an extensive investigation and optimisation of different methods and options, as
described in \cite{geisselsoeder:phdARXIV}. Since the resulting algorithm is
highly complex and not targeted at a specific signal configuration, however, it
is impossible to explore whether it is optimal in the full phase space of all
possible analysis methods.

Compared to the standard ANTARES point-source searches \cite{antares:ps}, the
multiscale method has been estimated to require a typically 20--100\% higher
neutrino flux in order to make an equally significant discovery for a point
source, depending on the assumed source spectrum. However, these previous
searches all target specific hypotheses of neutrino production and are not
designed to detect unexpected sources or distributions of sources.

\subsection{The search grid}

\begin{figure}[htb]
\centering
\includegraphics[width=\mySphereWidth\textwidth]{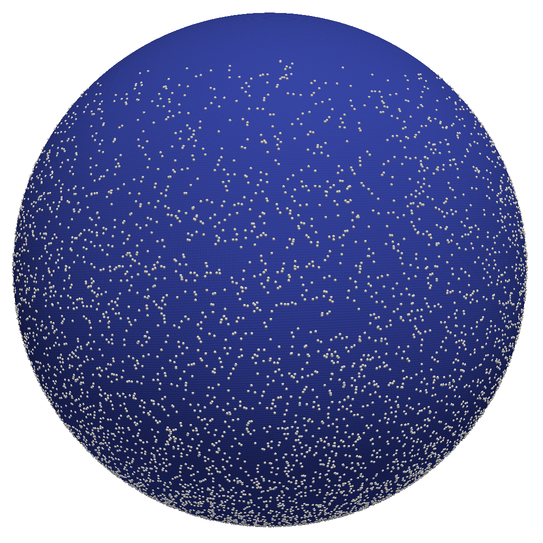}
\caption{%
A spherical grid in equatorial coordinates (in blue) with 12000 randomly
generated events and two point-like sources (in white). The grid points are
rendered with a radius of about $0.5^{\circ}$, hence they overlap and form a
closed sphere. Only the hemisphere of the three dimensional sphere facing the
observer is visible in this near-side general perspective projection. View on
the equator (declination of $0^{\circ}$).}
\label{fig:sourceSetup1}
\end{figure}

\begin{figure}[htb]
\centering
\includegraphics[width=\mySphereWidth\textwidth]{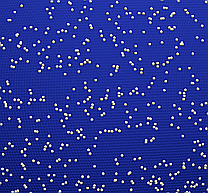}
\caption{%
A zoom to the centre of Figure~\ref{fig:sourceSetup1} where the grid points
become visible.}
\label{fig:sourceSetup2}
\end{figure}

\begin{figure}[htb]
\centering
\includegraphics[width=\mySphereWidth\textwidth]{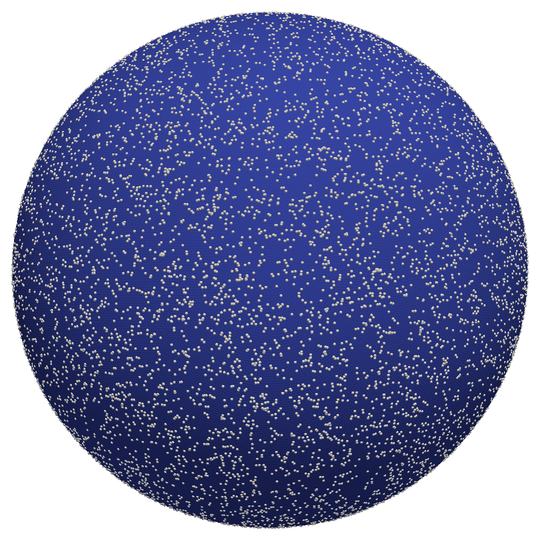}
\caption{
The same setup as in Figure~\ref{fig:sourceSetup1}, view from below on the south
pole (declination of $-90^{\circ}$). This setup contains the artificially added
events as shown more clearly in Figure~\ref{fig:sourceSetup4}. All following
spheres are oriented as in this Figure.}
\label{fig:sourceSetup3}
\end{figure}

\begin{figure}[htb]
\centering
\includegraphics[width=\mySphereWidth\textwidth]{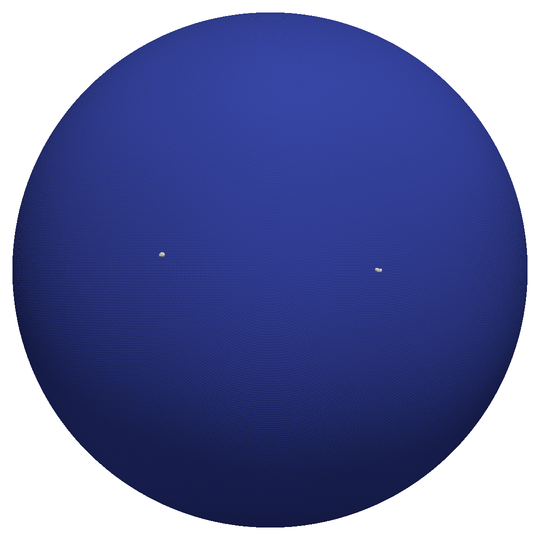}
\caption{
The same setup and orientation as Figure~\ref{fig:sourceSetup3}, but displayed
without the random events. The remaining white points are the events of the two
added point sources.}
\label{fig:sourceSetup4}
\end{figure}

In the first step a spherical grid is defined that allows for describing the sky
map in terms of sets of integers assigned to the grid points (see subsequent
steps). The distance of adjacent grid points is chosen to be $0.5^\circ$,
commensurate with the angular resolution of reconstructed neutrino-induced muon
tracks in ANTARES (see Fig.~\ref{fig:selectfit}). The grid consists of 165016
grid points. Figures~\ref{fig:sourceSetup1}, \ref{fig:sourceSetup2} and
\ref{fig:sourceSetup3} show the spherical grid, with grid points in blue and
12000 randomly generated neutrino events in white. This number of neutrinos is
close to the expectation for the data analysis and they are distributed
according to the visibility of ANTARES. In order to better illustrate the steps
of the analysis method, two artificial point sources with 12 and 18~events,
respectively, have been added at a declination of $-70^{\circ}$, see
Figures~\ref{fig:sourceSetup3} and \ref{fig:sourceSetup4}.

\subsection{Counting}

\begin{figure}[htb]
\centering
\includegraphics[width=0.5\textwidth]{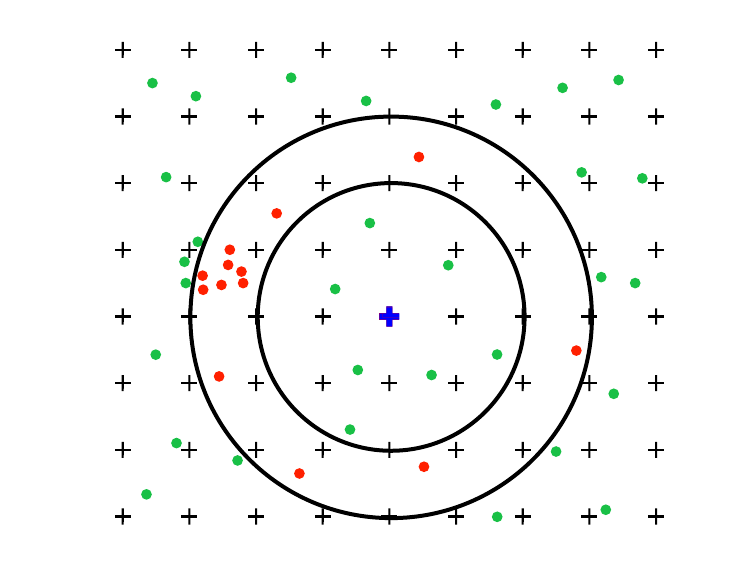}
\caption{%
Scheme of the neutrino counting. Crosses mark the grid points with a distance of
0.5$^{\circ}$ between them. Green and red dots correspond to neutrinos. The blue
cross is the grid point that is being evaluated. The search scale (here
1.0$^{\circ}$ to 1.5$^{\circ}$) is defined by the black circles. Neutrinos which
are counted with the current search scale around the search point are shown in
red. The results of the evaluation of this scale at the blue grid point is 13.}
\label{fig:searchRingScheme}
\end{figure}

\begin{figure}[p]
\centering
\includegraphics[width=\mySphereWidth\textwidth]{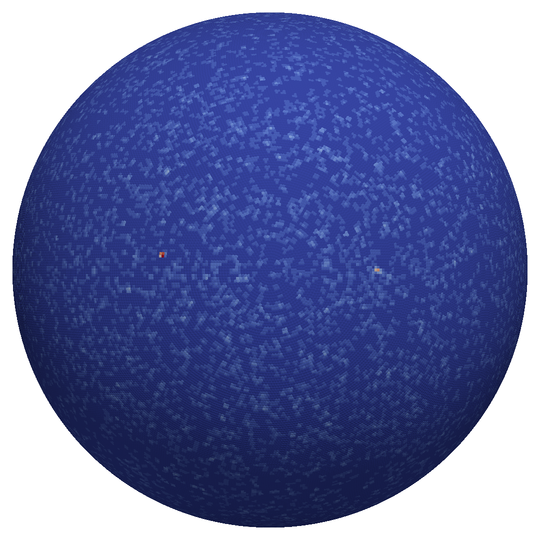}\\
\includegraphics[width=\mySphereWidth\textwidth]{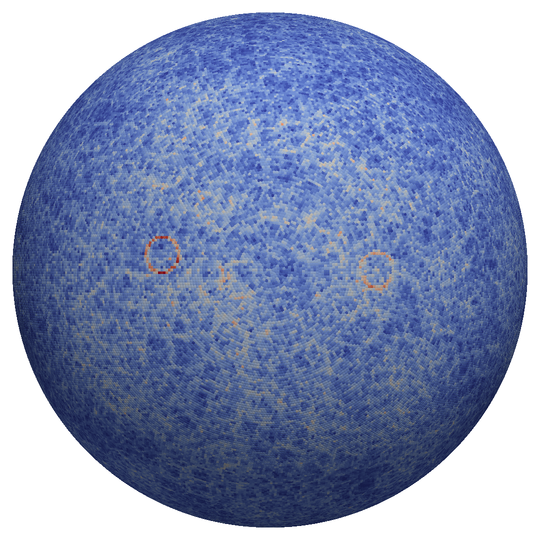}\\
\includegraphics[width=\mySphereWidth\textwidth]{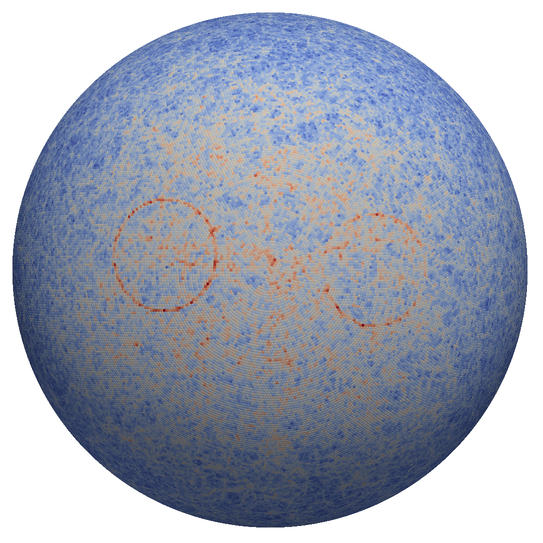}
\caption{%
The spherical search grid after counting the events presented in
Fig.~\ref{fig:sourceSetup3} in rings between 0$^{\circ}$ and 0.5$^{\circ}$
(top), between 3$^{\circ}$ and 3.5$^{\circ}$ (middle and between 10$^{\circ}$
and 10.5$^{\circ}$ (bottom) around each grid point. The colour code is
readjusted from top to bottom to the range of values present on each sphere.}
\label{fig:counting}
\end{figure}

In this step, a set of integers is determined for each grid point by counting
the neutrino events located in rings of different radii around that point; these
integers will be analysed in the following using Poissonian statistics. The
inner ring radius is referred to as the {\it search scale}, while the outer
radius is $0.5^\circ$ larger. The search evaluates 180 scales, i.e.\ 180
different regions with increasing angular radius from 0$^{\circ}$ up to
90$^{\circ}$ in steps of 0.5$^{\circ}$. This choice yields sensitivity to source
regions up to an extension of twice the maximal search scale, i.e.\ $180^\circ$. 
The counting is performed for each grid point and for each search scale. The
results are stored on an independent spherical grid for each search scale. A
visualisation of the counting scheme is depicted in
Figure~\ref{fig:searchRingScheme}. For three scales the results for the example
from Figure~\ref{fig:sourceSetup3} are shown in Figure~\ref{fig:counting}. Note
that, unless stated otherwise, in these and all following similar figures, the
colour code is readjusted between the different scales to match the full range
of values present on each sphere.

\subsection{Poisson probabilities}
\label{sec:poisson}

\begin{figure}[p]
\centering
\includegraphics[width=\mySphereWidth\textwidth]{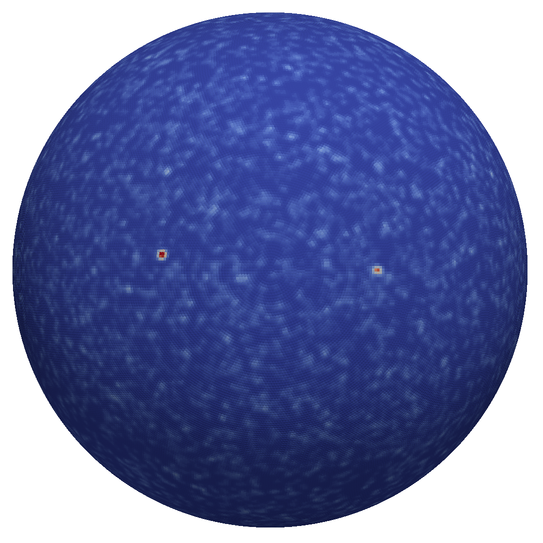}\\
\includegraphics[width=\mySphereWidth\textwidth]{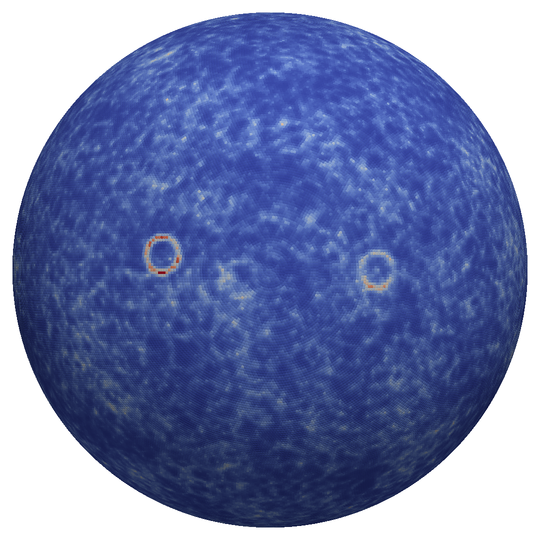}\\
\includegraphics[width=\mySphereWidth\textwidth]{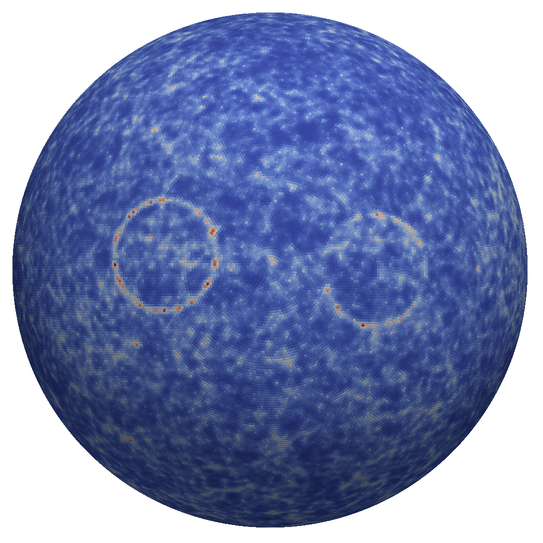}
\caption{%
Colour coded $R$ values as defined in Section~\ref{sec:poisson} derived from the
event counts shown in Fig.~\ref{fig:counting} for the scales 0$^{\circ}$ to
0.5$^{\circ}$ (top), 3$^{\circ}$ to 3.5$^{\circ}$ (middle), and 10$^{\circ}$ to
10.5$^{\circ}$ (bottom).}
\label{fig:countingToPoisson}
\end{figure}

Poissonian probabilities $P(N,\lambda)$ are now used to infer a measure for the
probability that the event numbers $N$ from the previous step exceed the
background expectation. The expected mean number $\lambda$ for each grid point
is estimated by pseudo-experiments using scrambled events. The scrambling is
achieved by randomly shuffling the event times before converting from local to
equatorial coordinates. This results in randomised distributions that preserve
the characteristics of the actual data taking, for instance the distribution of
the declinations or the efficiency of the data taking versus time. From the
$P(N, \lambda)$, the parameter
\begin{equation}
\label{equ:poissonAlter}
Q = \log_{10}\left(\frac{1}{P(x \ge N, \lambda)}\right) = -\log_{10} (P(x \ge N, \lambda) )
\end{equation}
is calculated. Subsequently, a low-pass filter\footnote{Implemented as
normalised box filter (averaging the values of eight neighbouring grid points
and the grid point itself)} is applied to the $Q$ values on the spheres to
reduce the influence of statistical fluctuations. These smoothed $Q$ values are
called $R$ and serve as input for the next step. The search spheres after these
computations are shown in Figure~\ref{fig:countingToPoisson}.

\subsection{Segmentation I}
\label{sec:segmentation1}

\begin{figure}[htb]
\centering
\includegraphics[width=0.48\textwidth]{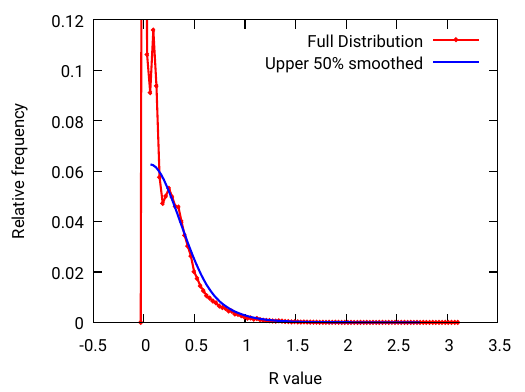}
\hfill
\includegraphics[width=0.48\textwidth]{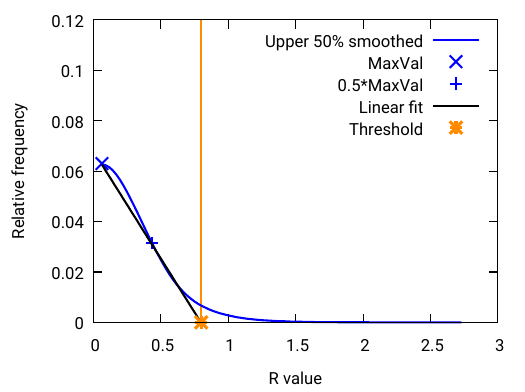}
\caption{%
Left: The red histogram shows the distribution of $R$ values for one scale
(0$^{\circ}$ to 0.5$^{\circ}$ in this example). Starting from the median value,
the histogram is smoothed by Gaussian smearing to obtain the blue curve. Right:
From the blue curve in the left panel the threshold $\theta$ is determined by
the zero-crossing of the line through the point with the maximum y-value on the
blue curve and the point with half this y-value.}
\label{fig:threshold}
\end{figure}

The next step aims to focus on potentially relevant information and to remove
background fluctuations. In general, this task, well known in the field of
computer vision, is called segmentation. Here it is performed using a simple
threshold $\theta$. Values below the threshold, e.g.\ less pronounced
fluctuations, are set to zero. The threshold $\theta$ is derived and applied for
each search scale independently, based on the histogram of the $R$ values
calculated for this scale. Figure~\ref{fig:threshold} illustrates the details
how the threshold is obtained. While multiple other options to determine a
threshold from the histogram would be possible, no method could be identified to
be significantly better than the others.

\begin{figure}[p]
\centering
\includegraphics[width=\mySphereWidth\textwidth]{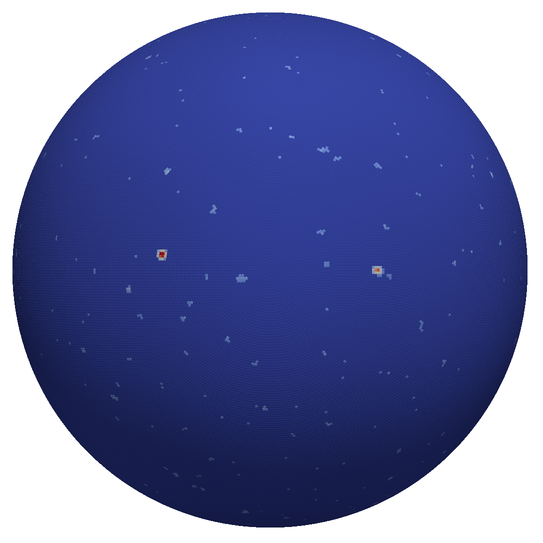}\\
\includegraphics[width=\mySphereWidth\textwidth]{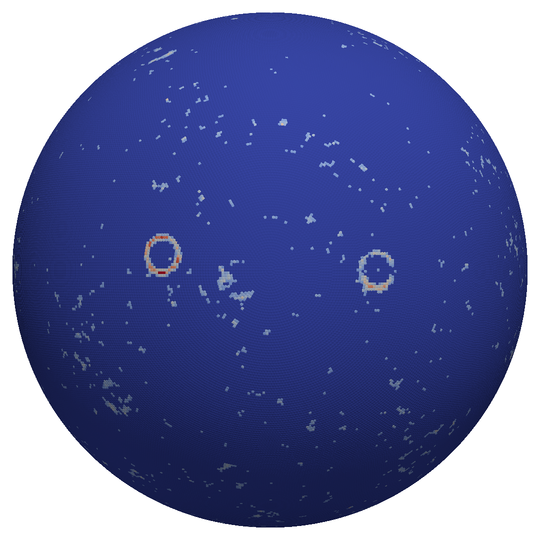}\\
\includegraphics[width=\mySphereWidth\textwidth]{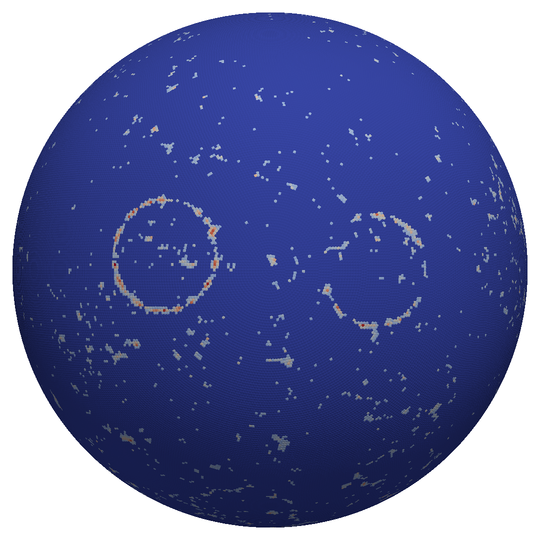}
\caption{%
$S$ maps after applying the segmentation step to the data shown in
Fig.~\ref{fig:countingToPoisson} according to Equation~\ref{equ:segmentation1}
for the scales 0$^{\circ}$ to 0.5$^{\circ}$ (top), 3$^{\circ}$ to 3.5$^{\circ}$
(middle), and 10$^{\circ}$ to 10.5$^{\circ}$ (bottom).}
\label{fig:PoissonToSegment}
\end{figure}

The outcome of this step at a grid point $p$ is given by: 
\begin{equation}
    S_p = \left\{\begin{array}{ll} 
    R_p, & R_p \ge \theta \\
    0, & R_p < \theta
    \end{array}\right. 
    \label{equ:segmentation1}
\end{equation}
The resulting $S$ maps after the segmentations are shown in
Figure~\ref{fig:PoissonToSegment}.

\subsection{Remapping}

\begin{figure}[p]
\centering
\includegraphics[width=\mySphereWidth\textwidth]{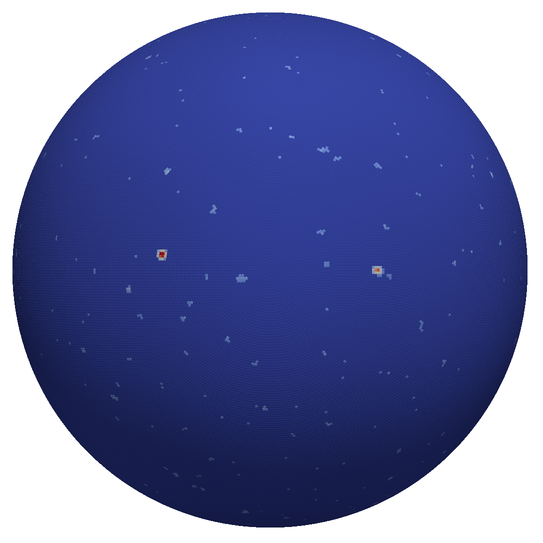}\\
\includegraphics[width=\mySphereWidth\textwidth]{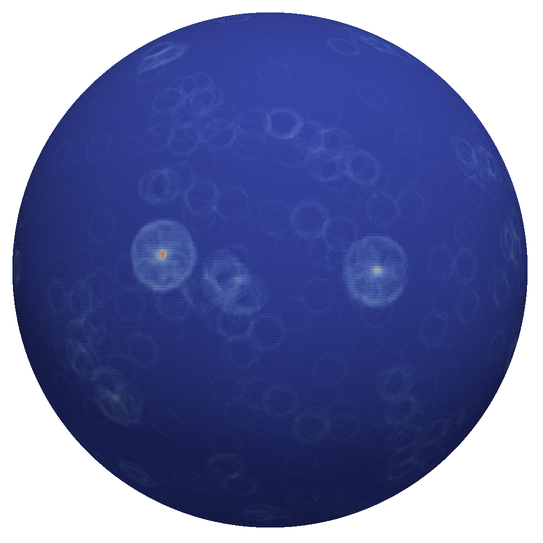}\\
\includegraphics[width=\mySphereWidth\textwidth]{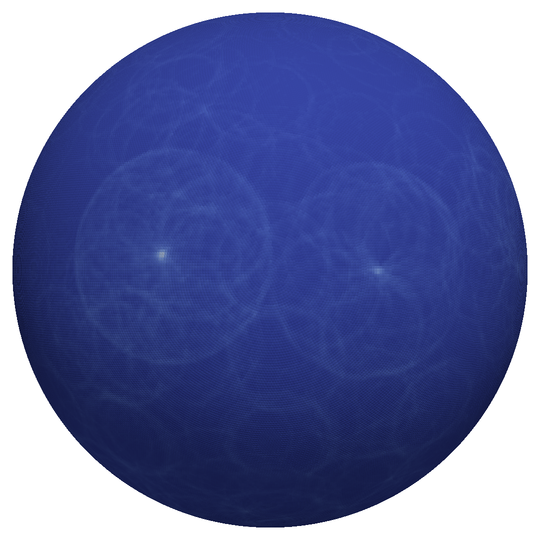}
\caption{%
Values of $T$ after the remapping step applied to the data shown in
Fig.~\ref{fig:countingToPoisson} according to Equation~\ref{equ:remap}, for the
scales 0$^{\circ}$ to 0.5$^{\circ}$ (top) 3$^{\circ}$ to 3.5$^{\circ}$ (middle),
and 10$^{\circ}$ to 10.5$^{\circ}$ (bottom). Note that the colour code is the
same for all three pictures.}
\label{fig:SegmentToRemap}
\end{figure}

The $S$ maps clearly show indications of the signal, but in a re-localised way --
for example, point sources are mapped to circles with the radius corresponding
to the search scale. The next step is to reconstruct the original location of
the neutrinos that caused the remaining high values of $S$ on the spheres. Since
the values for a given search scale $d$ originate from neutrinos that are
between $d$ and $d+0.5^\circ$ away from the grid points where the values are
stored, the following approach is used: let $\text{Set}_p$ be the set of $N_p$
grid points with a distance between $d$ and $d+0.5^\circ$ around a grid point
$p$ and let $q$ be an element (a grid point) of this set. The value $T_{p}$ for
grid point $p$ is calculated by averaging over the values $S_q$ of all grid
points in $\text{Set}_p$:
\begin{equation}
T_p = \sum_{q \in\text{Set}_p} \frac{S_q}{N_q}\,.
\label{equ:remap}
\end{equation}
This step maps the information back to all potential origins, meaning all grid
points where the neutrinos contributing to the value could have been located. 
Results of these computations are shown in Figure~\ref{fig:SegmentToRemap}. A
higher density in the original neutrino distribution is now encoded in the
overlapping pattern of the remapped circles as seen in the middle and bottom
pictures of Figure~\ref{fig:SegmentToRemap}. Note that the halos around the
original source positions are unavoidable artifacts of this method, which reduce
its sensitivity for point-source searches but less so for extended sources.

\clearpage

\subsection{Summation}

\begin{figure}[htb]
 \centering
   \includegraphics[width=\mySphereWidth\textwidth]{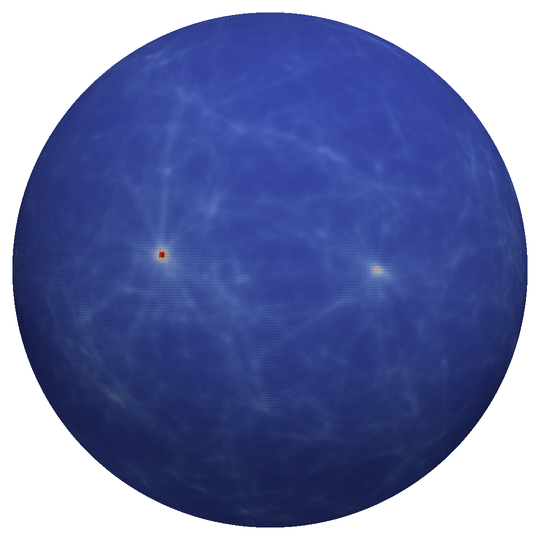}
\caption{%
Values of $U$, given by Equation~\ref{equ:sum}.}
\label{fig:sumSphere}
\end{figure}

So far, all calculations have been made independently for each search scale. In
order to derive information on size and topology of potential signal regions
without imposing a distance or size scale, the results for the different search
scales need to be combined. Having evaluated multiple approaches to exploit the
information available in the multitude of scales, a simple sum of the maps
assigned to all scales of a given spherical grid was found to yield the most
robust evaluation. Since the influence of random fluctuations is high for the
smallest scale, it is not included in this sum. The input for the final steps is
calculated as:
\begin{equation}
U_p = \sum_{i=2}^{180} T_{p,i}\,,
\label{equ:sum}
\end{equation}
where the index $i$ denotes the distance scale, given by $d = (i-1) \cdot
0.5^{\circ}$. This index has been omitted in the previous equations as all
computations have been restricted to the same distance scale. The result of the
summation according to Equation~\ref{equ:sum}, applied to the full data set and
all scales, can be seen in Figure~\ref{fig:sumSphere}.

\subsection{Segmentation II}

\begin{figure}[p]
\centering
\includegraphics[width=\mySphereWidth\textwidth]{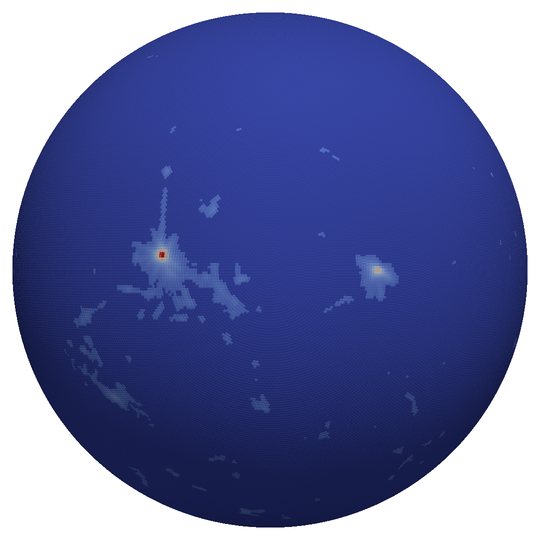}\\
\includegraphics[width=\mySphereWidth\textwidth]{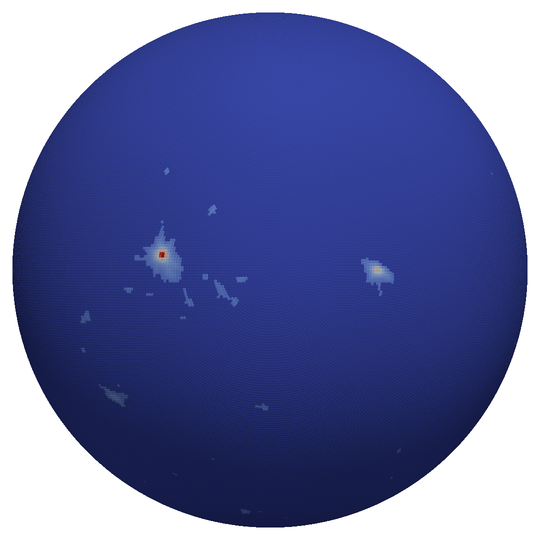}\\
\includegraphics[width=\mySphereWidth\textwidth]{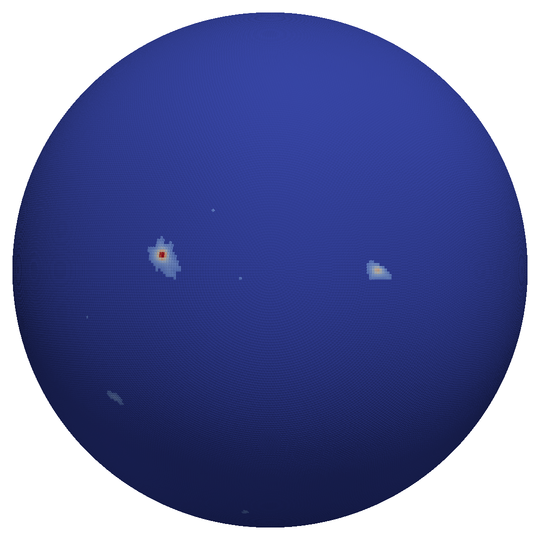}
\caption{%
$V$ values after the segmentation as given by Equation~\ref{equ:segmentation2}
for $\beta = 0.75$ (top), $\beta = 0.95$ (middle), and $\beta = 1.11$ (bottom).}
\label{fig:segmentationComparison}
\end{figure}

A further segmentation is performed to isolate potential signal regions from
background fluctuations. The same procedure which was used before for the
histograms of $R$ values (see~Section~\ref{sec:segmentation1}) is used here for
the single histogram of $U$ values. In this step, however, different thresholds
$\theta_{\beta}$ are used, obtained by scaling the difference between the
minimum non-zero value found on the sphere, $U_{\mathrm{min}}$, and $\theta$,
the threshold computed as in the previous segmentations, by a factor $\beta \in
\mathbb{R}$:
\begin{equation}
\label{equ:thresholdAlpha}
\theta_{\beta} = U_{\mathrm{min}} + \beta(\theta-U_{\mathrm{min}})\,.
\end{equation}
The variable threshold $\theta_\beta$ serves to adjust the sensitivity of the
segmentation step. The result of the segmentation using $\beta$ for each grid
point $p$ is given by:
\begin{equation}
    V_p = \left\{\begin{array}{ll} 
    U_p, & U_p \ge \theta_{\beta} \\
    0, & U_p < \theta_{\beta}
    \end{array}\right. 
    \label{equ:segmentation2}
\end{equation} 
Lower values for $\beta$ result in lower thresholds $\theta_{\beta}$, allowing
more extended structures to be found, while higher values only preserve the high
peaks, favouring smaller morphologies. The results are finally filtered using a
two-dimensional median filter\footnote{The value at a grid point is replaced by
the median value of the eight neighbouring grid points and itself.} to suppress
potential artefacts (e.g. single grid points). The effect of different
thresholds for the segmentation can be seen in
Figure~\ref{fig:segmentationComparison}. Choosing the value of $\beta$ is the
first step in this analysis where an explicit bias for any source property is
included. Based on an evaluation of a variety of simulated sources, two values
are used: $\beta = 0.75$ and $\beta = 1.11$. These values are chosen to be more
sensitive to extended than to small (or point-like) structures. This choice is
motivated by the objective to perform a search that complements the previous,
location-independent ANTARES searches for point-like sources \cite{antares:ps}.

\subsection{Significance}
\label{sec:significance}

A possible signal of the analysis will manifest itself as a connected group of
grid points, all with values above $\theta_{\beta}$; such an object is called a
``cluster''.

The final step is to distinguish potentially significant clusters from random
accumulations. Since the exact size, shape, position and composition of a
cluster is highly unlikely to be reproduced using pseudo-experiments, more
generic metrics must be used to evaluate the significance of a cluster. Many
metrics have been designed and tested on a multitude of simulated sources
\cite{geisselsoeder:phdARXIV}, each with different sensitivities to different
sources. No single metric can be maximally sensitive to all potential sources. 
Motivated by the focus on extended sources, the metric ``clustersize'', given by
the number $N_c$ of grid points in a cluster, has been chosen, performing best
for extended source topologies. The quantity $N_c$ is mainly correlated wit the
size of the signal region. Some information on the density of neutrinos within a
cluster is nevertheless retained, because the average value of $U$ in a given
cluster, and hence the cluster size, grows with the neutrino density.

\begin{figure}[htb]
 \centering
   \includegraphics[width=0.5\textwidth]{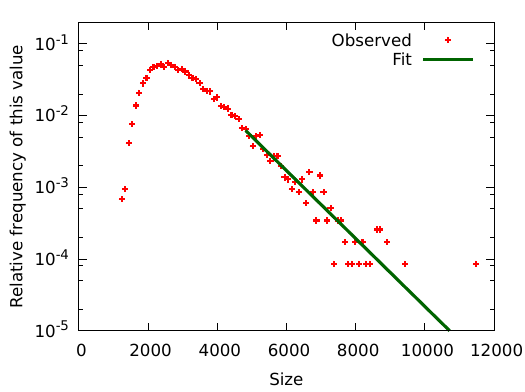}
\caption{%
The distribution of the sizes of clusters from pseudo-experiments and a fit to
the tail of the distribution.}
\label{fig:significanceSize}
\end{figure}

The significance for a cluster is derived from pseudo-experiments with scrambled
events as explained in Section \ref{sec:poisson}. For each threshold
$\theta_{\beta}$ the distribution obtained from the pseudo-experiments has to be
treated independently, as e.g.\ the distribution of the sizes of clusters
depends on the used threshold. A pre-trial p-value and hence a pre-trial
significance is computed for each observed cluster using the corresponding
distribution. The distribution of the values obtained for the metric
``clustersize'' (in grid points) for a threshold using $\beta = 0.75$ is shown
in Figure~\ref{fig:significanceSize}. An exponential function $a \cdot e^{bx}$
with free parameters $a$ and $b$ is fitted to the tail of the distribution. This
function is used instead of the histogram if the significance determination
would be limited by the available statistics. For each pseudo-experiment, only
the highest pre-trial significance of any cluster from any threshold is
considered the result of this pseudo-experiment. The post-trial significance is
then obtained by a comparison of the measured pre-trial significance of an
observed cluster with the distribution obtained from pseudo-experiments.

\subsection{Application to extended sources and general considerations}

\begin{figure}[htb]
\centering
\includegraphics[width=\mySphereWidth\textwidth]{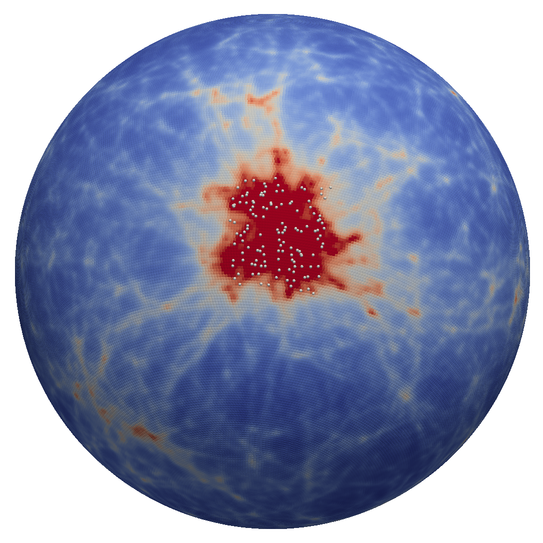}
\caption{%
An illustrative example for the behaviour of the method for an extended source,
20$^\circ$ by 20$^\circ$. The additional 120 source neutrinos are depicted in
white. The 12000 background neutrinos are not rendered. Colour indicates the $U$
values after the summation, corresponding to Figure~\ref{fig:sumSphere}. }
\label{fig:otherSources1}
\end{figure}

\begin{figure}[htb]
\centering
\includegraphics[width=\mySphereWidth\textwidth]{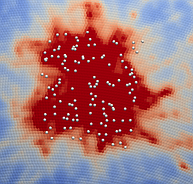}
\caption{%
A zoom to the source region of Figure~\ref{fig:otherSources1}.}
\label{fig:otherSources2}
\end{figure}

\begin{figure}[htb]
\centering
\includegraphics[width=\mySphereWidth\textwidth]{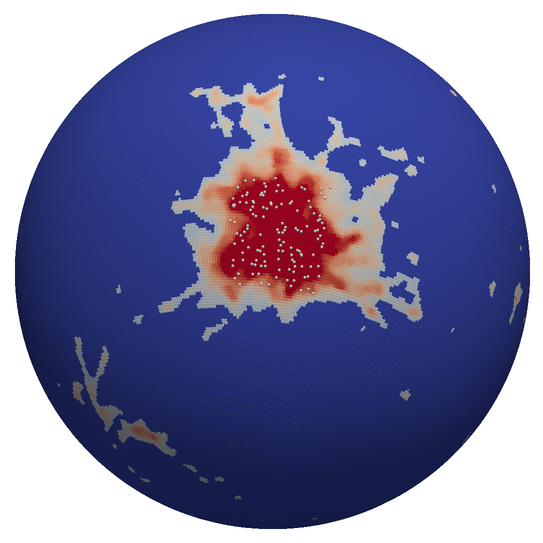}
\hfill
\includegraphics[width=\mySphereWidth\textwidth]{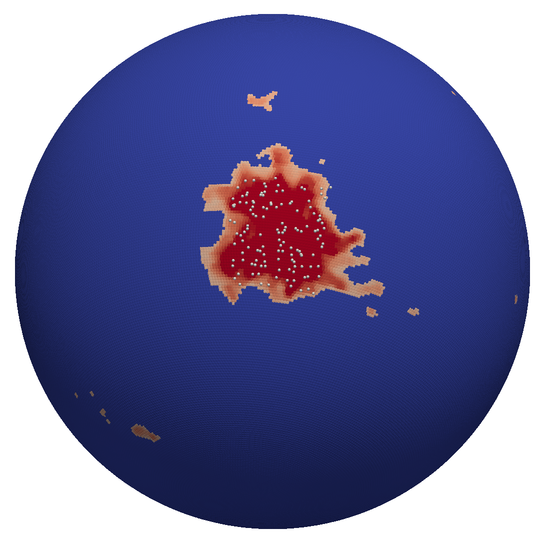}
\caption{%
The results for the setup in Figure \ref{fig:otherSources1} after a segmentation
using $\beta = 0.75$ (left) and $\beta = 1.11$ (right), respectively,
corresponding to the $V$ values in the top panel of
Figure~\ref{fig:segmentationComparison}.}
\label{fig:otherSources34}
\end{figure}

To illustrate the behaviour of the method for a simulated extended source, the
configuration shown in Figures~\ref{fig:otherSources1} and
\ref{fig:otherSources2} has been investigated. As shown in
Figure~\ref{fig:otherSources34}, the location and size of the simulated signal
are approximated reasonably well. For the result obtained with the lower $\beta$
value, however, additional filaments extend from the actual shape to regions
where random accumulations of background neutrinos occurred.

The sensitivity of this method for deviations from a homogeneous spatial
distribution is different compared to previously used algorithms. For instance,
tests with the same simulated pseudo-experiments showed that the two-point
correlation analysis \cite{antares:2pt} is more sensitive to scenarios where
many faint sources with the same extension are distributed evenly throughout the
sky. On the other hand, the multiscale search described here is more sensitive
in scenarios that include a spatial clustering of these faint sources. It also
benefits more from the presence of one or more dominant sources.

The additional information reported in \cite{geisselsoeder:phdARXIV}
(pp.~95--98) verify an important aspect of the method: It only requires an
increase in the number of excess events by a factor of six for an increase of
the signal region from $(2\,\text{degrees})^2$ to $(20\,\text{degrees})^2$. 
Since a factor of ten would be expected for a constant signal-to-noise ratio, it
can be concluded that the multiscale search is most sensitive to large-scale
clusterings and thus has a unique sensitivity characteristic, clearly different
from most other search methods.

More examples of sources, also for asymmetric shapes, and more details on the
quantitative effect of neighbouring sources can be found in
\cite{geisselsoeder:phdARXIV}.

While the true nature of the sources of high-energy cosmic neutrinos is still
unknown, this algorithm offers a high sensitivity for a wide range of possible
scenarios, including ones in which many faint sources dominate the observed
diffuse flux \cite{Aartsen:2017}.

\FloatBarrier

\section{Results}

This analysis was performed following a data blinding concept. That means that
the development of all methods presented in
Sections~\ref{sec:dataSelectionAndReco} and~\ref{sec:multiscale} and the
optimisation of all cuts had been completely finished before the recorded data
sample was processed. The small data sample that had been used to verify the
methods was excluded from the final sample.

\subsection{ANTARES}

\begin{figure}[ht]
\begin{center}
\includegraphics[width=\mySkymapWidth\textwidth]{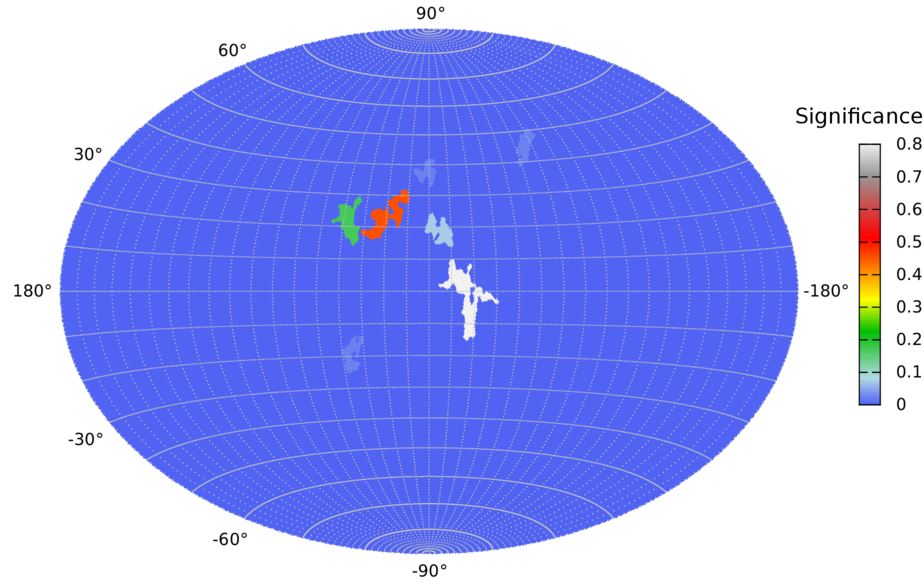} \\
\includegraphics[width=\mySkymapWidth\textwidth]{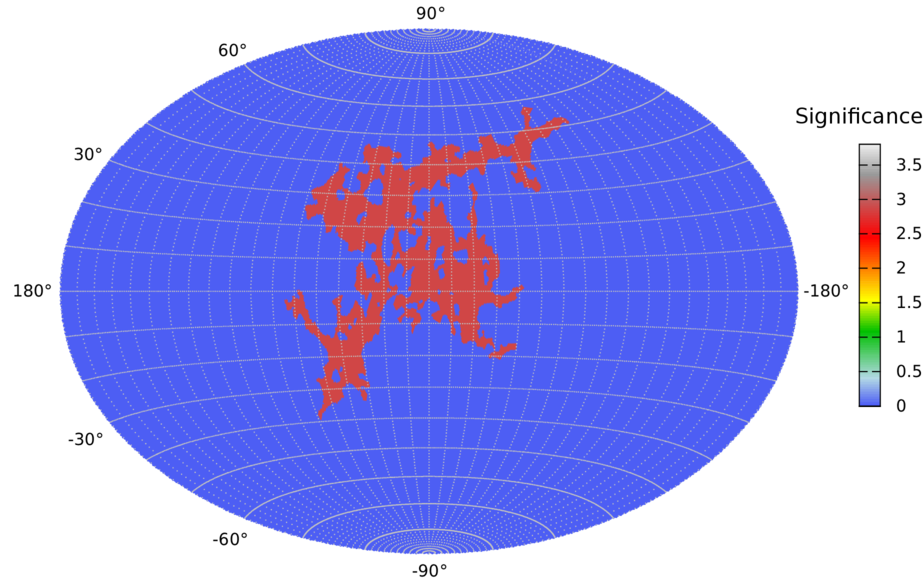}
\end{center}
\caption{%
The result of the multiscale search for ANTARES data. The skymap is a
Hammer-Aitoff projection of the resulting sphere in galactic coordinates. The
colour code of the clusters indicates the post-trial significance in units of
$\sigma$. Top: using $\beta = 1.11$ for segmentation. Bottom: Using $\beta =
0.75$. }
\label{fig:resultANTARES}
\end{figure}

Using the methods described in Sections \ref{sec:updown} and
\ref{sec:selectfit}, the evaluation of the ANTARES data from 2007 to 2012
results in 13283 candidates for charged current muon neutrino events, with an
expected number of background neutrinos of $13078\pm362$ (statistical error)
from interactions of cosmic rays in the atmosphere. This corresponds to a
background expectation of $0.634$ neutrino candidates$/\text{degree}^2$. The
multiscale search method described in Section \ref{sec:multiscale} yields the
result shown in Figure~\ref{fig:resultANTARES}. Using the higher segmentation
threshold $\beta = 1.11$, no cluster with a significance above $0.8\sigma$ has
been found. With $\beta = 0.75$ a very large structure is found. After
accounting for all known systematic effects, the large structure, called ``the
cluster'' from here on, has a post-trial significance of $2.5\sigma$. The
investigation of the systematic effects can be found in
\cite{geisselsoeder:phdARXIV}. The size of the cluster is 13312 connected grid
points, equivalent to about 3328\,degree$^{2}$ or 8\% of the sky.

The region contains about 200~events in addition to the about 2100~events
expected from background. This corresponds to an excess of $0.06\pm0.015$
neutrino candidates$/\text{degree}^2$. These numbers are derived assuming no
subclustering within the cluster. Any subclustering (e.g.\ by embedded smaller
clusters) would result in the same significance for a smaller average neutrino
excess density. This localized event excess would not have been noticed by the
two-point correlation analysis \cite{antares:2pt} nor the point source analysis
and does not violate the existing bounds and observations from ANTARES and
Icecube on the astrophysical diffuse flux.

\begin{figure}[ht]
\begin{center}
\includegraphics[width=\mySkymapWidth\textwidth]{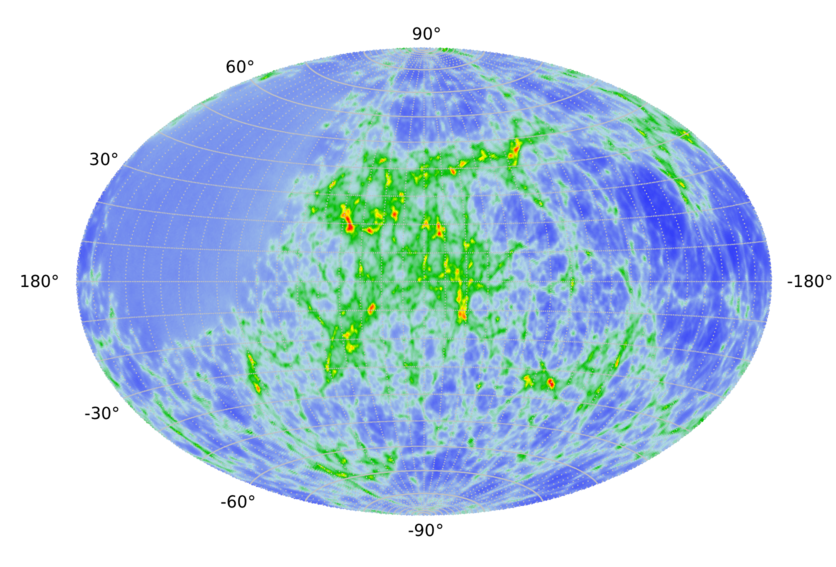} \\
\includegraphics[width=\mySkymapWidth\textwidth]{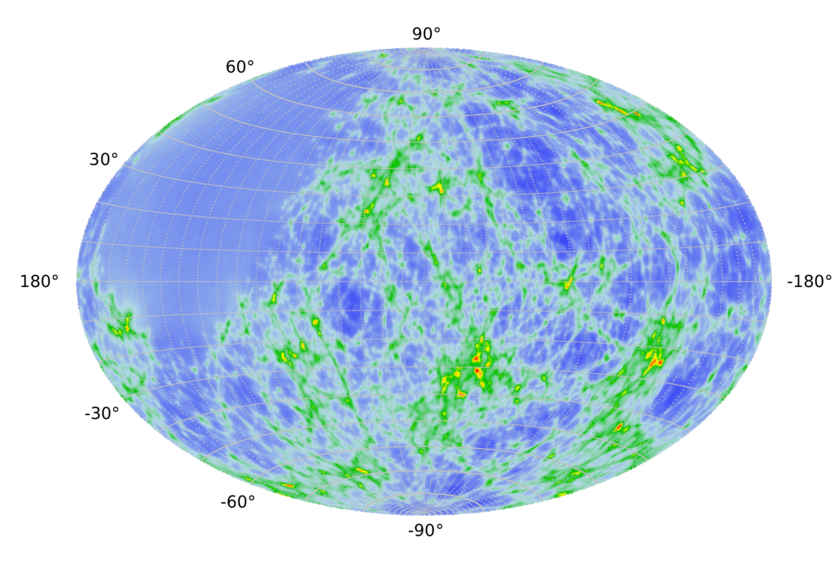}
\end{center}
\caption{%
Top: Sum of all evaluated scales (corresponding to Figure~\ref{fig:sumSphere},
not directly to the neutrino flux) resulting from ANTARES data, in galactic
coordinates and before segmentations. As only upgoing events are used in this
analysis, the field of view of ANTARES does not cover the whole sky. This
results in the large, homogeneously blue area in the upper left of the skymap. 
Bottom: For comparison, an example for a random dataset with the same colour
code as the ANTARES results. The observed maximum values of $U$, see
Equation~\ref{equ:sum}, are similar, while the clustering of the
overfluctuations is more pronounced in the recorded data.}
\label{fig:detailedANTARES}
\end{figure}

Even if not significant on its own, this structure constitutes an interesting
feature in the data which is worth further studies. It can be noted that the
cluster contains the Galactic Centre, which is located in the centre of the
presented skymaps in galactic coordinates. More details on the inner structure
of these clusters can be seen in Figure~\ref{fig:detailedANTARES}, which shows
the result of the summation of all scales before the segmentations. It should be
noted that with the limited available statistics, random fluctuations do
influence the results.

\subsection{IceCube IC40}
\label{sec:IC40}

In order to perform an independent cross check of the result obtained using
ANTARES data, the publicly available IC40 dataset \cite{ice:ic40c} published by
the IceCube Collaboration has also been analysed. This analysis searches
specifically for an excess in the area of the large cluster found in ANTARES
data. To achieve this, it only considers clusters found in the IC40 data that
overlap with the ANTARES cluster by at least 51\% of their area. The value of
51\% is chosen here because this means that they are more related to the cluster
from ANTARES than to other regions. On the other hand, requiring an overlap of
close to 100\% would exclude e.g. clusters that extend beyond the shape of the
ANTARES cluster.

Since the requirement for overlap implicitly confines the size of the cluster,
the clustersize in grid points, $N_c$, is no reasonable metric to assess the
significance of a cluster in this evaluation. Therefore the metric that gave the
second best results in the investigations introduced in Section
\ref{sec:significance} has been used. It is the mean value of the $\sqrt{N_c}$
grid points\footnote{Rounded to the nearest integer number} with the highest
values within a cluster.

\begin{figure}[ht]
\begin{center}
\includegraphics[width=\mySkymapWidth\textwidth]{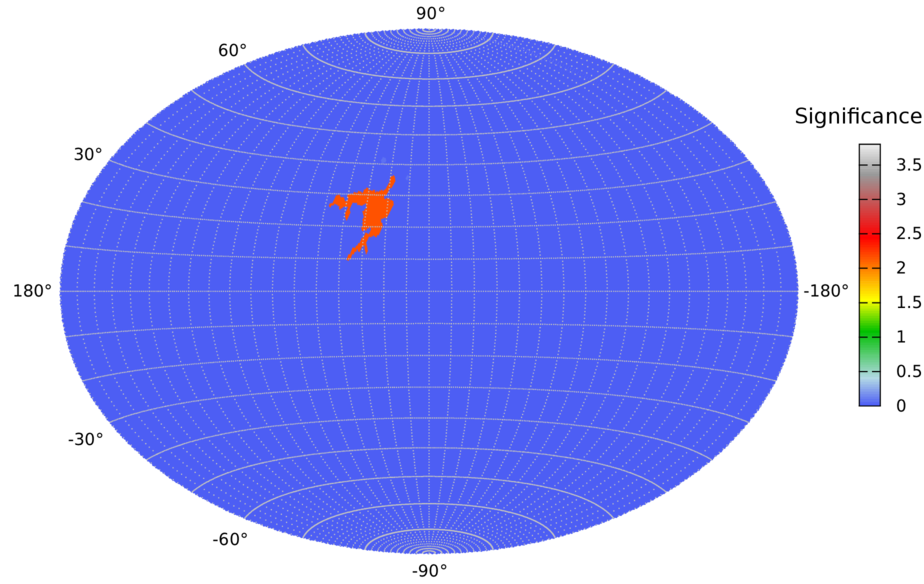} \\
\includegraphics[width=\mySkymapWidth\textwidth]{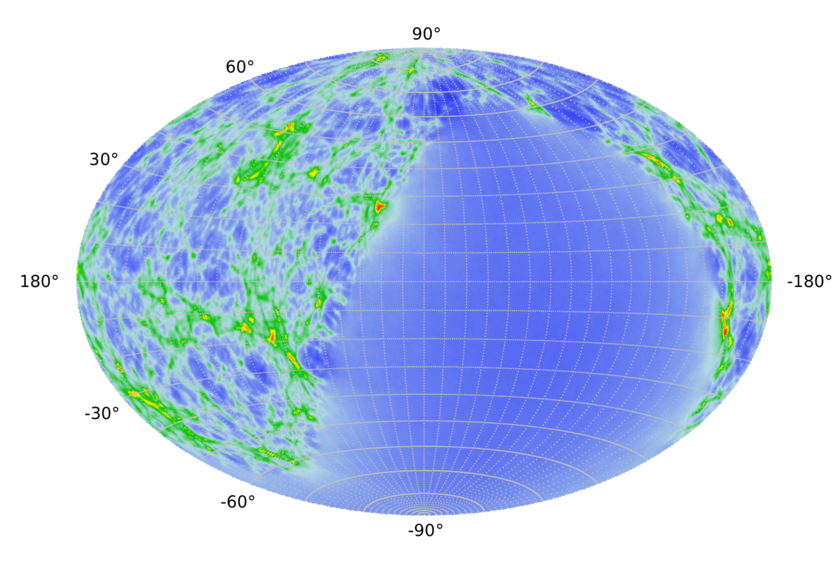}
\end{center}
\caption{%
Top: The result on public IC40 data with segmentation using $\beta = 0.75$ in
galactic coordinates. Bottom: The detailed structure of $S$-values behind the
result on IC40 data before the segmentation.}
\label{fig:resultIC40}
\end{figure}

The result obtained by this adapted search on the IC40 dataset is shown in
Figure~\ref{fig:resultIC40}. A cluster is found within the expected shape with a
post-trial significance of 2.1$\sigma$. The position of the found cluster
overlaps with the search template by 78\% of its size.

The observation that not all features in the skymaps in Figures
\ref{fig:detailedANTARES} and \ref{fig:resultIC40} match exactly is to be
expected, as ANTARES has its highest sensitivity at lower energies compared to
IceCube. Moreover, random fluctuations of atmospheric neutrinos do influence the
results.

\section{Conclusions}

This paper presents three original and unpublished tools that have been used for
a search for cosmic neutrino sources of arbitrary location and morphology by
detecting the most pronounced deviation from the background-only expectation. 
The study is fully data-driven, without relying on any model for neutrino
emission and input derived from simulations.

The first tool (see Sect.~\ref{sec:updown}) is a multivariate classification
technique used to increasing the available statistics of ANTARES data. The
second (see Sect.~\ref{sec:selectfit}) enhances the directional reconstruction
of charged-current muon neutrino candidates. The third tool is a
model-independent multiscale method (described in Sect.~\ref{sec:multiscale})
that represents the core of this study.

Due to the fact that the method is independent of neutrino emission models or
assumptions on the topology of emission regions, this analysis is not intended
to discover potential sources with the highest achievable significance, nor to
analyse the properties of a candidate source region. Instead, it aims at
indicating the clustering in the data least expected from the background-only
hypothesis and thus to initiate more specific investigations.

Applied to ANTARES data recorded between 2007 and 2012, this analysis found a
large structure with a post-trial significance of 2.5$\sigma$. This result is
consistent with a random fluctuation of the background of atmospheric neutrinos. 
Using this method to analyse public data from IceCube resulted in an excess
located within the overlap between the cluster from the ANTARES data and the
field of view of IceCube. This observation has a significance of 2.1$\sigma$.

Even though a general, model-independent analysis cannot be as sensitive as a
dedicated search due to the high trial factors, this method provides a good way
to become aware of the most prominent and even unforeseen structures in data and
can be regarded as a trigger for more specific investigations.
Despite the high trial-factor, it can outperform standard point-source and
two-point correlation analyses in a scenario of unknown extended structures.

\FloatBarrier

\section*{Acknowledgments}

The authors acknowledge the financial support of the funding agencies:
Centre National de la Recherche Scientifique (CNRS), Commissariat \`a
l'\'ener\-gie atomique et aux \'energies alternatives (CEA),
Commission Europ\'eenne (FEDER fund and Marie Curie Program),
Institut Universitaire de France (IUF), IdEx program and UnivEarthS
Labex program at Sorbonne Paris Cit\'e (ANR-10-LABX-0023 and
ANR-11-IDEX-0005-02), Labex OCEVU (ANR-11-LABX-0060) and the
A*MIDEX project (ANR-11-IDEX-0001-02),
R\'egion \^Ile-de-France (DIM-ACAV), R\'egion
Alsace (contrat CPER), R\'egion Provence-Alpes-C\^ote d'Azur,
D\'e\-par\-tement du Var and Ville de La
Seyne-sur-Mer, France;
Bundesministerium f\"ur Bildung und Forschung
(BMBF), Germany; 
Istituto Nazionale di Fisica Nucleare (INFN), Italy;
Stichting voor Fundamenteel Onderzoek der Materie (FOM), Nederlandse
organisatie voor Wetenschappelijk Onderzoek (NWO), the Netherlands;
Council of the President of the Russian Federation for young
scientists and leading scientific schools supporting grants, Russia;
National Authority for Scientific Research (ANCS), Romania;
Mi\-nis\-te\-rio de Econom\'{\i}a y Competitividad (MINECO):
Plan Estatal de Investigaci\'{o}n (refs. FPA2015-65150-C3-1-P, -2-P and -3-P, (MINECO/FEDER)), 
Severo Ochoa Centre of Excellence and MultiDark Consolider (MINECO), and Prometeo and Grisol\'{i}a programs (Generalitat
Valenciana), Spain; 
Ministry of Higher Education, Scientific Research and Professional Training, Morocco.
We also acknowledge the technical support of Ifremer, AIM and Foselev Marine
for the sea operation and the CC-IN2P3 for the computing facilities.

\bibliography{myReferences}

\end{document}